\documentclass{iaus}
\usepackage{graphicx}

\title[Infrared Observations of Massive SFRs with Masers] 
{Masers and the Massive Star Formation Process: New Insights Through Infrared Observations}

\author[De Buizer et al.]   
{James M. De Buizer$^1$, James T. Radomski$^{1,2}$, Charles M. Telesco$^2$, \break \and Robert K. Pi\~{n}a$^3$}

\affiliation{$^1$Gemini Observatory, Casilla 603, La Serena, Chile \break email: jdebuizer@gemini.edu \\[\affilskip]
$^2$Department of Astronomy, University of Florida, Gainesville, FL, 32601 USA\\[\affilskip]
$^3$Photon Research Associates, Inc., 5720 Oberlin Drive, San Diego, CA 92121 USA}

\pubyear{2005}
\volume{227}  
\pagerange{1-6}
\date{May 6}
\jname{Massive Star Birth: A Crossroads of Astrophysics}
\editors{R. Cesaroni, E. Churchwell, M. Felli, \& C.M. Walmsley, eds.}
\begin{document}

\maketitle

\begin{abstract}
Our mid-infrared and near-infrared surveys over the last five years have helped to strengthen and clarify the relationships between water, methanol, and OH masers and the star formation process. Our surveys show that maser emission seems to be more closely associated with mid-infrared emission than cm radio continuum emission from UC HII regions. We find that masers of all molecular species surveyed trace a wide variety of phenomena and show a proclivity for linear distributions. The vast majority of these linear distributions can be explained by outflows or shocks, and in general do not appear to trace circumstellar disks as was previously thought. Some water and methanol masers that are not associated with radio continuum emission appear to trace infrared-bright hot cores, the earliest observable stage of massive stellar life before the onset of a UC HII region.

\end{abstract}

\firstsection 
\section{Introduction}

Originally masers were found in regions of radio continuum and thus thought to trace massive star formation sites. Soon after, as advances in spatial resolution of radio observations improved, it was discovered that masers are sometimes associated with ultra-compact HII (UC HII) regions of ionized gas from young massive stars, but sometimes masers were found offset from these radio continuum regions. If the masers are found in star-forming regions but are not all coincident and tracing massive stars, what processes or phenomena associated with star formation do these masers trace?

To better understand the maser phenomena and the relationship between masers and the star formation process we must know the locations of all the associated young stellar sources in the star-forming regions containing the masers in question. Traditional radio continuum observations alone will only reveal \emph{some} stellar sources in the field. However, massive stars form in clusters with variety of masses and states of evolution. Therefore we need an independent tracer of star formation.

Mid-infrared emission traces the hot dust close to the stellar sources while penetrating the cooler obscuring dust of the molecular cloud in which it resides. Also, using 4-10m class telescopes we can achieve resolutions comparable to interferometric radio continuum observations of the ionized gas from the UC HII regions.

\section{The Infrared Surveys}


Over the last 5 years we have been conducting surveys of massive star-forming regions in the mid-infrared (De Buizer et al. 2000; De Buizer et al. 2003; De Buizer et al. 2005), and to a less extent in the near-infrared (De Buizer 2003), to learn more about the relationship between masers and young stars. We present here some of what we have learned from these surveys and follow-up observations.

A general result from these surveys is that the mid-infrared detection rate towards sites of methanol maser emission is 67\%. This is very comparable to our mid-infrared detection rate rate towards sites with water masers of 77\%. Some care must be taken with interpreting these results since it has been found recently that many of the methanol maser selected sites have water maser emission, and vice-versa.

Another interesting general result from the water maser survey is that the median separation between centers of water maser emission and mid-infrared sources was only $\sim$8700 AU. Compared to the median separation of water maser centers to radio UC HII regions ($\sim$18800 AU; Hofner \& Churchwell 1996) this seems to show that masers are more closely associated with mid-infrared sources than UC HII regions (a similar comparison could not be made for the methanol maser survey because of a larger error in the absolute astrometry of those mid-infrared observations).

\section{Spatial Linearity of Maser Spots}

The first survey to spatially resolve methanol maser emission (Norris et al. 1993) showed that the emission was found to come in groups of individual masers spots. Norris et al. (1993) pointed out that it appeared that these spot distributions tended to be linearly distributed on the sky in a large portion of their survey sample. Because this seemed to be a statistically significant portion of population, it was thought that methanol masers must therefore be special and trace a peculiar phenomenon. Early on, shock and outflows were ruled out as the cause of these linear distributions, and the idea that these masers were tracing edge-on disks around young massive stars was forwarded. This claim has generated a lot of interest because the idea that massive stars form via accretion is still in debate, and as a consequence it is not known if they should indeed have circumstellar accretion disks.

One result that came out of our water maser survey had nothing to do with the mid-infrared observations themselves. Interestingly, further inspection of our sample of water maser sources (chosen solely by right ascension), showed the water maser spots within groups are linearly distributed in 38\% of the cases. This matches the rate at which we see methanol masers linearly distributed (37\% in Walsh et al. 1998, 38\% in Phillips et al. 1998). Also, for our water maser selected sample we had data for OH masers in the regions as well (Forster and Caswell 1989). We found in our sample that OH maser groups are even more prone to linear distributions at the rate of 50\%.

Many of these linear water (and OH) maser distributions are found associated with outflows, shocks, and other phenomena in our surveys and in the observations of other authors. It seems that methanol masers are not special for displaying linearity in their maser groups, and that perhaps, as we shall see in the next section, masers tracing disks may be the exception instead of the rule.

\section{Masers as a Tracer for Circumstellar Disks}

Though maser linearity is a compelling argument in favor of the disk hypothesis, the main problem is that there is very little supporting evidence from observations at other wavelengths. One example claiming to directly detect the disk associated with masers was G339.88-1.26 (Stecklum \& Kaufl 1998), seen as an elongated mid-infrared source at the same location and elongated at the same position angle as the linearly distributed methanol masers. However follow-up mid-infrared observations at higher resolution resolved the ``disk'' into three separate sources (De Buizer et al. 2002).

\begin{figure}
\includegraphics[width=5.2in]{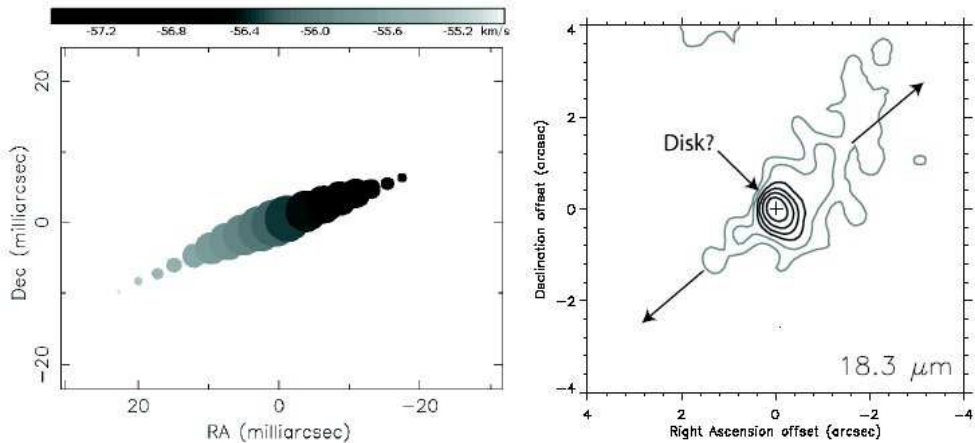}
  \caption{The disk and outflow of NGC 7538 IRS 1.
    (\textit{a}) Methanol masers (dots) are linearly distributed with a velocity gradient along the distribution (size of dots proportional to their relative flux densities). Pestalozzi et al. (2004) modeled the masers as being in a circumstellar disk;
    (\textit{b}) However there is an outflow at a similar angle as the maser elongation as seen in CO (arrows), and the diffuse mid-infrared emission (gray) appears to be tracing the outflow cavity. The bright core in the mid-infrared (black) is elongated perpendicular to this outflow and may be the disk collimating the outflow. The location of the center of the line of masers (cross) coincides with the peak of this mid-infrared source (from De Buizer \& Minier 2005).}\label{fig:contour}
\end{figure}

There are also a few examples where outflows are found perpendicular to linear maser distributions. This is thought to be evidence supporting the disk scenario since disks are thought to be outflow collimators. One such example is the young solar mass star NGC 2071 IRS1 discovered by Torrelles et al. (1998) to have a linear distribution of water masers perpendicular to a radio jet. A similar (but less convincing) example for a case involving a high-mass young stellar source is G192.16-3.82, where the water masers appear to be linearly distributed and arranged perpendicular to an outflow seen in CO (Shepherd \& Kurtz 1999). It must be stressed that it is not certain if the masers are indeed tracing a disk in either of the two above cases.

The most recent linear distribution of methanol masers that is gaining a lot of attention is NGC 7538 IRS 1, and is presently thought to be the best case for masers tracing a disk (Fig. 1a). Not only do the masers form a tight linear structure, they have a velocity pattern along the maser spot distribution that can be fit by a Keplerian disk of 750 AU in radius rotating around a 30 solar mass star (Pestalozzi et al. 2004). In order to investigate this hypothesis further, De Buizer \& Minier (2005) performed high resolution mid-infrared observations of the NGC 7538 IRS 1 region. In conjunction with CO outflow observations of the region (Davis et al. 1998), the mid-infrared observations show that the linear methanol maser distribution is parallel to the outflow coming from IRS 1. Also IRS 1 is elongated in its mid-infrared dust emission, but at an angle perpendicular to the linear distribution of methanol masers and the outflow (Fig. 1b). Therefore, if there is indeed a dusty circumstellar disk around IRS 1 it may be elongated perpendicular to the linear maser distribution.

In summary, the momentum in favor of the disk hypothesis for masers seems to be waning. However, as is seen for NGC 7538 IRS 1 and as we shall see in the next section, linear distributions of methanol masers are being found {\it parallel} to outflows. Therefore there is growing evidence that suggests that there may be a more direct relationship between methanol masers and outflows from young stars.

\section{The Relationship between Masers and Outflow}

Of all the maser species, water masers have been known for the longest time to have an association with outflows from young stars. In fact, water masers are not only associated with the outflows from high-mass young stellar sources, but young low-mass stars as well. Usually these water masers have seemingly random distributions on the sky, however occasionally there have been sources seen with linear, X-shaped, or V-shaped maser distributions that are associated with outflows (Norris et al. 1993; Minier et al. 2000).

\begin{figure}
\begin{center}
\includegraphics[width=2.5in]{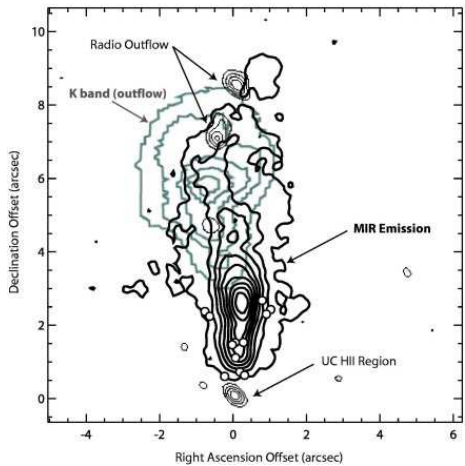}
  \caption{The outflow from G35.20-0.74. The outflow was established by observations in radio continuum emission (thin black; Gibb et al. 2003) and in the near-infrared at K (thick gray; Fuller et al. 2003). We can clearly see the outflow cavity of this source in the mid-infrared at 10 $\mu$m (thick black) emanating from the UC HII region at the bottom of the figure. Water masers (circles) are in a V-shape and trace the walls of the outflow as seen in mid-infrared emission.}\label{fig:contour}
\end{center}
\end{figure}

In order to test the disk hypothesis for linearly distributed methanol masers, De Buizer (2003) searched for outflow signatures perpendicular to the maser position angles, under the assumption that the disks would collimate outflows in a direction normal to the disks. De Buizer (2003) performed imaging of these sources in the the 2.122 $\mu$m H$_{2}$ line, which is a near-infrared indicator for shock-excited emission from outflows. Of the 15 sources where the H$_{2}$ was believed to be outflow related, it was found that 12 sources (80\%) had H$_2$ emission found in regions dominantly \emph{parallel} to the linear methanol maser distributions, opposite of what was expected for the disk scenario. Instead, what these observations may be indicating is a direct link between methanol masers and the outflow process itself. The methanol masers, like some water masers, may be created by shocks and heating on the working surfaces of the outflow cavities carved out of the molecular core where their stellar host resides. However, unlike water masers, it is not generally believed that methanol masers are shock excited, but instead radiationally excited by mid-infrared photons (Cragg et al. 2005).

New observations of sources like G35.20-0.74 (De Buizer et al., in prep) are showing that outflow cavities from massive stars can be highly luminous in the mid-infrared. For G35.20-0.74, there is a V-shaped distribution of water masers tracing the outflow cavity walls seen in the mid-infrared (Fig. 2). These water masers are likely shock-excited by the outflow through this cavity. However, for other similarly embedded high-mass stellar sources it appears that the heating of these cavity walls by the central star can provide enough mid-infrared emission to pump and excite methanol maser emission.

\section{The Association of Masers and Hot Cores}

Some masers and maser groups are found offset and isolated from UC HII regions in regions of massive star formation. Though these masers are not coincident with radio continuum emission, many are found to be associated with sources seen in sub-mm or mid-infrared continuum emission, and/or molecular line emission. Since masers are thought to trace high-mass stars in general, and these isolated masers reside in regions of high-mass star formation (as evidenced by the presence of nearby UC HII regions), it is thought that these sources are the youngest stages of massive stellar life before the onset of a UC HII region. These types of sources have come to be called hot cores.

\begin{figure}
\includegraphics[width=5.2in]{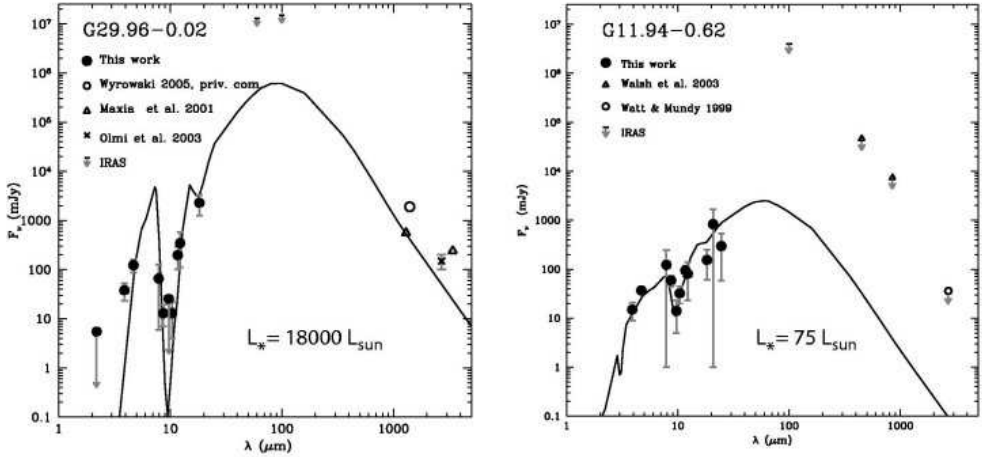}
  \caption{SEDs for two hot core candidates associated with water maser emission from De Buizer, Osorio, \& Calvet (2005).
    (\textit{a}) The known hot core in G29.96-0.02 has a high luminosity confirming its massive nature,
    (\textit{b}) however the hot core candidate in G11.94-0.62, though observationally similar to the G29.96-0.02 hot core, appears to be underluminous for a massive star, and is likely a young intermediate mass protostar.}\label{fig:contour}
\end{figure}

In mid-infrared surveys of De Buizer et al. (2003) and De Buizer et al. (2005), several water maser groups were found to be isolated from radio continuum emission, but were coincident with bright mid-infrared sources. Some of these sources were known hot cores already found by their molecular line emission. These observations were followed up by observations at many mid-infrared wavelengths using narrowband filters (De Buizer, Osorio, \& Calvet 2005). When combined with millimeter and sub-millimeter continuum measurements, the SEDs for these sources could be modeled and estimates could be made for several physical parameters, including luminosity (Fig. 3). It was found that there are sources that have similar observational properties described above for hot cores, but are in fact lower (intermediate) mass young stellar objects (Fig. 3b).

Therefore, although masers can be associated with young high-mass stars before the onset of UC HII regions, some stellar sources associated with masers have no radio continuum emission simply because they are not massive enough to ionize their surroundings. The mid-infrared derived luminosities from De Buizer et al. (2000) and De Buizer et al. (2005), though crudely derived lower limits, show a range of source luminosities associated with water, methanol, and hydroxyl masers from that range from spectral types A to O. It seems plausible that many of the sources with A and B mid-infrared derived spectral types are genuine intermediate-mass stellar sources.

\section{Summary}

Mid-infrared observations are a great complement to radio continuum observations of star-forming regions. Together they reveal a majority of the stellar sources in the vicinity of the masers and allow a better interpretation of the relationship between the masers and the star formation regions than either wavelength alone.

Masers of \emph{all} molecular species appear to form in groups and often have linear distributions. This does not appear to be a property specific to methanol masers, as was originally thought. There are a limited number of sources where it appears that these linearly distributed masers may be tracing circumstellar disks, however the vast majority of linearly distributed masers are not associated with disks. Indications from outflow observations towards regions that have linearly distributed masers are that the masers are more likely to be associated with outflows.

Furthermore, it appears that some water and methanol masers may trace hot cores, the earliest stages of massive star formation before the onset of a UC HII region. However, at least in the case of water masers, some sources have been observed to have similar observations properties to massive hot cores but are in fact intermediate mass young stellar objects.

\end{document}